\documentclass[prx,preprint,superscriptaddress,amsmath,amssymb,floatfix]{revtex4-1}

\usepackage{array}
\usepackage{graphicx}
\usepackage{bm}
\usepackage[usenames]{color}
\usepackage[utf8]{inputenc}
\usepackage{color,soul}

\newcommand{\beq}{\begin{equation}}
\newcommand{\eeq}{\end{equation}}
\newcommand{\beqa}{\begin{eqnarray}}
\newcommand{\eeqa}{\end{eqnarray}}

\begin{document}

\title{Mechanical analogue of a Majorana bound state}

\author{Chun-Wei Chen}
\thanks{These authors contributed equally}
\affiliation{Aeronautics and Astronautics, University of Washington, Seattle, WA 98195-2400, USA}
\author{Natalia Lera}
\thanks{These authors contributed equally}
\affiliation{Departamento de F\'isica de la Materia Condensada,
Universidad Aut\'onoma de Madrid, Madrid 28049, Spain,
Condensed Matter Physics Center (IFIMAC) and Instituto Nicol\'as Cabrera}
\author{Rajesh Chaunsali}
\affiliation{Aeronautics and Astronautics, University of Washington, Seattle, WA 98195-2400, USA}
\author{Daniel Torrent}
\affiliation{GROC, UJI, Institut de Noves Tecnologies de la Imatge (INIT), Universitat Jaume I, 12071, Castell\'o, Spain}
\author{Jose Vicente Alvarez}
\affiliation{Departamento de F\'isica de la Materia Condensada,
Universidad Aut\'onoma de Madrid, Madrid 28049, Spain,
Condensed Matter Physics Center (IFIMAC) and Instituto Nicol\'as Cabrera}
\author{Jinkyu Yang}
\email[Corresponding author.\\]{jkyang@aa.washington.edu}
\affiliation{Aeronautics and Astronautics, University of Washington, Seattle, WA 98195-2400, USA}
\author{Pablo San-Jose}
\affiliation{Materials Science Factory, Instituto de Ciencia de Materiales de Madrid (ICMM-CSIC), Sor Juana In\'es de la Cruz 3, 28049 Madrid, Spain}
\author{Johan Christensen}
\email[Corresponding author.\\]{johan.christensen@uc3m.es}
\affiliation{Department of Physics, Universidad Carlos III de Madrid, ES-28916 Legan\`es, Madrid, Spain}

\date{\today}


\maketitle  

\textbf{The discovery of topologically non-trivial electronic systems has opened a new age in condensed matter research. From topological insulators to topological superconductors and Weyl semimetals, it is now understood that some of the most remarkable and robust phases in electronic systems (e.g., Quantum Hall or Anomalous Quantum Hall) are the result of topological protection. These powerful ideas have recently begun to be explored also in bosonic systems. Topologically protected \emph{acoustic} \cite{yang2015topological,he2016acoustic,NatPhys.13.369,TopoSound}, \emph{mechanical} \cite{BertoldiGyro,VitelliGyro,SusstrunkScience,torrentLike,PhysRevX.8.031074} and \emph{optical} \cite{PhysRevLett.100.013904,NatMatKhanikaev,LingLu,Ma_2016,RechtsmanReview} edge states have been demonstrated in a number of systems that recreate the requisite topological conditions. Such states that propagate without backscattering could find important applications in communications and energy technologies. In this work we demonstrate the mechanical analogue of a topologically \emph{bound} state, a different class of non-propagating protected state that cannot be destroyed by local perturbations. These are well known in electronic systems, such as Majorana bound states \cite{Kitaev:PU01,Fu:PRL08} in topological superconductors, but remain largely unexplored in a bosonic setting. We implement topological binding by creating a Kekulé distortion vortex \cite{Kekule:ACP66} on a two-dimensional mechanical honeycomb superlattice.}

Topologically protected states emerge at generic topological defects of a gapped band structure. This general statement takes its simplest form in topological insulators~\cite{Hasan:RMP10}, electronic systems with a gap at the Fermi level that is topologically distinct from that of vacuum or conventional insulators. At the boundary between a topological insulator and vacuum, the topological index is forced to change and thus, the boundary is a topological defect from the electronic point of view.  It develops protected edge states that cannot be destroyed, as long as the basic symmetries of the system (time-reversal symmetry in this example) are preserved. In a deep sense, these topological states are a manifestation of bulk topology, and their emergence is sometimes dubbed the bulk-boundary correspondence principle.

Topological modes at an extended boundary are propagating, but this is not generic. Topological bound states are also possible, e.g., at the point-like boundaries between two one-dimensional (1D) topological insulators or superconductors~\cite{Jackiw:PRD76}. The 1D topological superconducting case is particularly striking. The topological states at its boundaries have zero energy, and are known as Majorana bound states (MBSs), due to their peculiar self-conjugate $\gamma=\gamma^\dagger$ nature (half-electron, half-hole). These states, invented by Ettore Majorana in 1937 in a very different context~\cite{Majorana:INC137} and studied by Kitaev in 2001 in superconductors~\cite{Kitaev:PU01}, have been proposed as a building block of practical quantum computers~\cite{Nayak:RMP08}, and are the subject of intense research currently~\cite{Elliott:RMP15,Aguado:RNC17}. MBSs are furthermore a realisation of the topological binding mechanism studied in the context of the 1D Dirac equation by Jackiw and Rebbi in 1976~\cite{Jackiw:PRD76}. 

\begin{figure}
   \centering
   \includegraphics[width=\columnwidth]{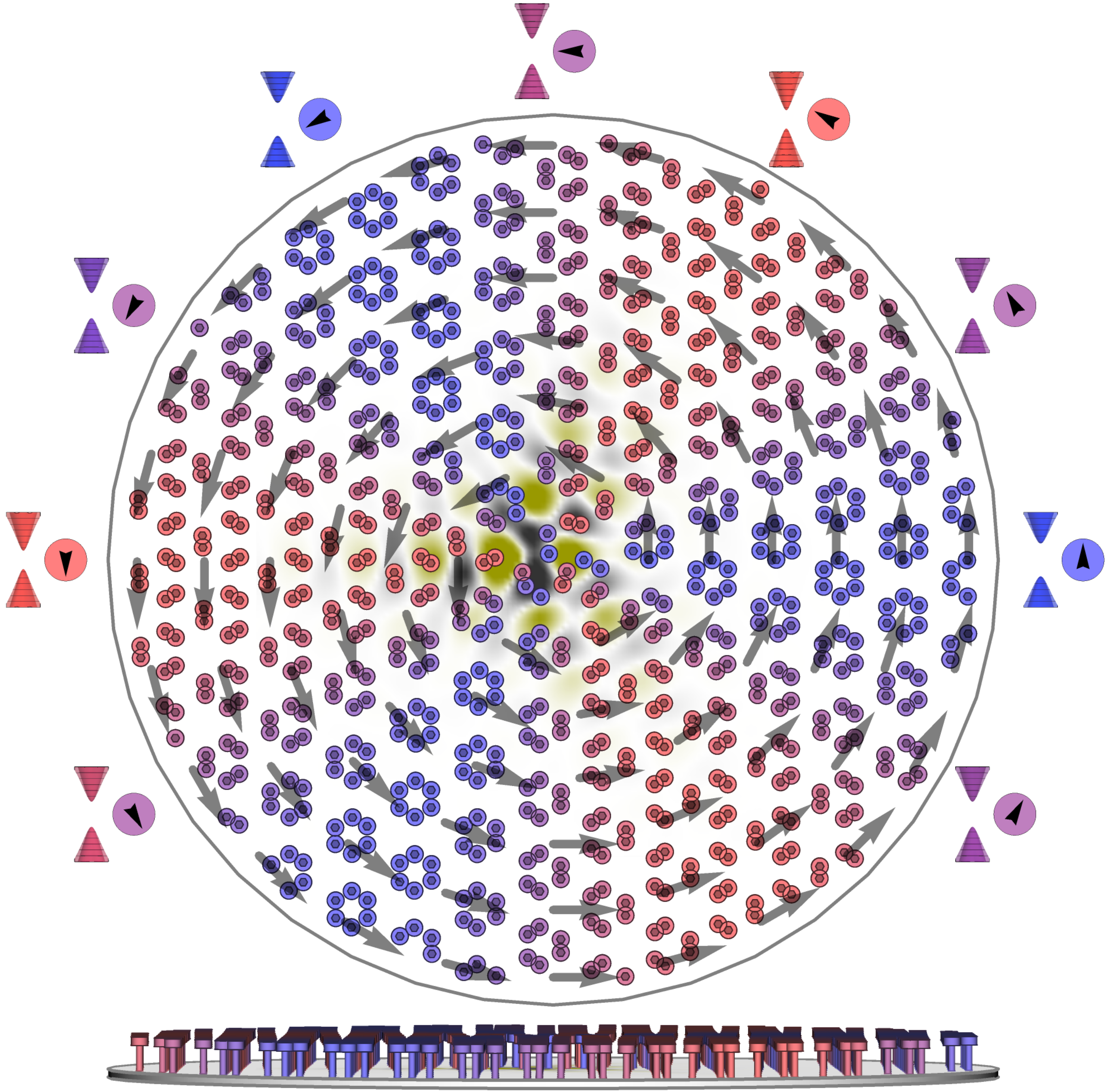}    
   \caption{\textbf{Kekul\'e distorted mechanical graphene}. A honeycomb lattice of bolts is attached to an oscillating plate. Shifting the position of bolts following a  Kekul\'e pattern gaps the modes. The Kekul\'e gap has an internal phase $\phi(\bm{r})$ in valley space (see bands in perimeter) that can create a vortex. A Kekul\'e vortex traps a topologically protected mode (in black and yellow), a mechanical analogue of a Majorana bound state.}
   \label{fig:sketch}
\end{figure}

A related work by Jackiw and Rossi demonstrated topological binding of states in the two-dimensional Dirac equation by means of vortices in a gap-opening field~\cite{Jackiw:NPB81}. Magnetic flux vortices in spinless, $p$-wave superconductors have been shown to be a  realisation of this proposal~\cite{Ivanov:PRL01, Nishida:PRB10}, with topologically trapped MBS emerging at the vortex core. The fundamental Jackiw-Rossi Hamiltonian takes the low-energy form of two Dirac valleys $s=\pm 1$ coupled by a gap-opening parameter $\Delta$,
\begin{equation}
H = \int d^2r 
\sum_{s=\pm}\psi^\dagger_s\left(\vec{\sigma}\cdot\vec{\partial}_r\right)\psi_s + \Delta\psi^\dagger_+\psi_{-} + \Delta^*\psi^\dagger_-\psi_{+}.
\end{equation}
The gap parameter can be position dependent and is in general complex, $\Delta(\bm{r})=|\Delta(\bm r)| e^{i\phi(\bm r)}$. A magnetic $n$-vortex is realized by a winding of $2\pi n$ of its phase $\phi$ around the vortex core (taken here at the origin $\bm r=0$), where continuity forces the modulus $|\Delta|$ to vanish. A standard model for a vortex of radius $\xi$ reads~\cite{Shore:PRL89}
\begin{equation}
\label{Delta}
\Delta(\bm{r}) = \Delta_0\tanh(r/\xi)\exp(in\phi_{\bm{r}}),
\end{equation}
where $\bm{r} = r(\cos\phi_{\bm{r}}, \sin\phi_{\bm{r}})$.
This vortex traps a Majorana zero energy state, confined to the core of the vortex. No perturbation to the system, other than merging of two vortices, can remove a MBS from zero energy. Key to this protection is the charge-conjugation symmetry between electrons and holes, which is also responsible for the particle-hole symmetry of the Bogoliubov--de Gennes spectrum of $H$.

In this work we demonstrate a mechanical analogue of a MBS topologically bound to a vortex. The field of topological mechanics recently emerged as a fertile ground to explore bosonic analogies in terms of elastic wave motions. Along this frontier, artificial gyroscopic lattices have shown to sustain robust chiral edge states by mimicking the quantum Hall effect \cite{BertoldiGyro,VitelliGyro}. Other phononic metamaterials have been constructed to enable mechanical versions of the quantum spin Hall effect \cite{SusstrunkScience,torrentLike,PhysRevX.8.031074} and the valley degree of freedom \cite{RuzzeneNJPhBN} for unidirectional phononic signal guiding.

\begin{figure*}
   \centering
   \includegraphics[width=0.9\textwidth]{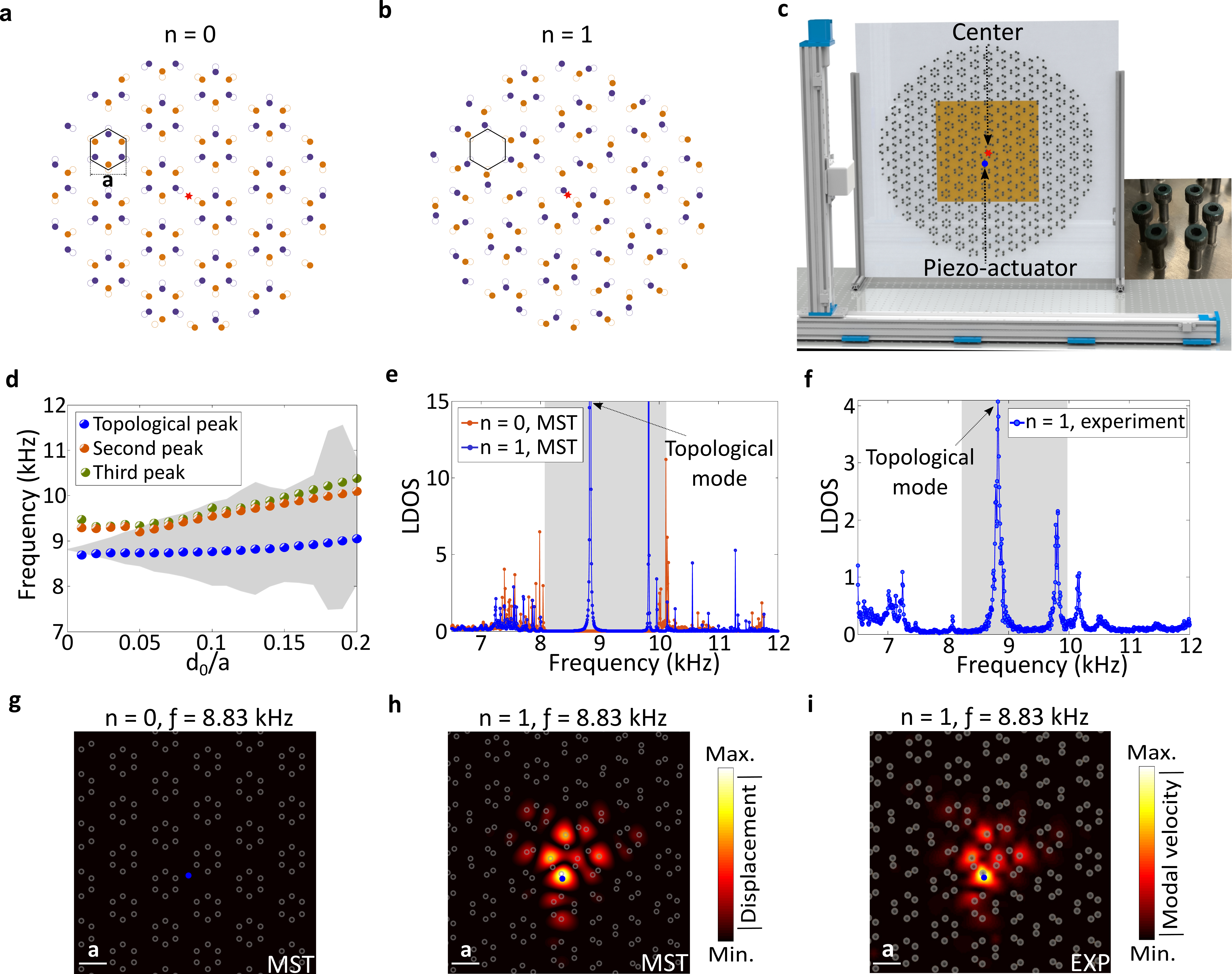}    
   \caption{\textbf{System configuration and the vortex bound states}. 
\textbf{a, b}, No-vortex ($n=0$) and vortex ($n=1$) patterns, respectively. The hollow circles containing two sub-lattices in golden and purple denote unperturbed ($d_0=0$) honeycomb configuration with lattice period $a$, whereas the filled circles denote the perturbed configuration as per Eq.~\ref{dr}. The red star denotes the origin. \textbf{c}, Experimental setup, in which a laser Doppler vibrometer scans the yellow area to measure the out-of-plane response of the $n=1$ vortex lattice. The pattern contains 1069 bolts mounted on a thin aluminium plate as shown in the inset. The blue dot represents the location where the piezoelectric actuator has been mounted. \textbf{d}, MST predictions of the bandgap opening (in grey) and the three localized states that become bound as one increases the perturbation strength $d_0/a$ for $n=1$. The topological peak converges to the Dirac point as the perturbation is reduced to zero. The other two peaks enter the continuum for small perturbations. From this point on the vortex amplitude is fixed to $d_0/a=0.15$. \textbf{e}, LDOS at the core, predicted by MST for the $n=0$ and $n=1$ structures. \textbf{f}, Measured LDOS verifying the presence of localized states inside the bandgap. \textbf{g}, Spatial profile of the out-of-plane displacements within the bandgap when the $n=0$ structure is excited at the blue point. \textbf{h, i}, Simulated displacements and measured power spectral density for $n=1$ at their corresponding frequencies.}
   \label{Fig2}
\end{figure*}

We constructed an artificial  honeycomb-like  lattice, i.e., \emph{distorted mechanical graphene}, by decorating a thin aluminum plate with steel bolts~\cite{torrent2013elastic,torrentLike, chaunsali2018experimental}. The pattern of bolts, rendered in Fig.~\ref{fig:sketch}, realises a periodic honeycomb lattice plus a distortion field. The former produces a spectrum analogous to graphene's, with two valleys and a Dirac-like dispersion around a specific Dirac frequency in each of them. As we will see in the following, this approach emulates electronic topological phenomena in our bosonic mechanical structure. The distortion field for an $n$-Kekulé-vortex takes the form of a position-dependent displacement $\delta\bm{r}(\bm{r})$ of the resonator (bolt) sites,
\begin{equation}
\label{dr}
\delta\bm{r}(\bm{r})=d(r)\left[\sin(\bm{K}\bm{r}+\phi(\bm{r})), \pm\cos(\bm{K}\bm{r}+\phi(\bm{r}))\right],
\end{equation}
where $\bm{K}=[4\pi/3a, 0]$ is the valley wavevector, $\pm$ correspond to the two sublattices, and $(d, \phi)$ play the role of the modulus and phase of $\Delta(\bm{r})$, respectively. For a constant $d=d_0$ and $\phi$, the lattice remains periodic (see Fig.~\ref{Fig2}a), but the perturbation triples the unit cell, which folds the two Dirac cones onto the $\Gamma$-point and opens a gap $\sim d_0/a$. The gap changes slightly as a function of the phase $\phi$, but never closes. In the rest of this work we focus on the $n=1$ vortex, defined by $d(r) = d_0 \tanh(r/\xi)$ and $\phi(\bm{r})=\phi_{\bm{r}}$, with $\xi=0.1a$ and $d_0=0.15 a$ specifically. The vortex lattice is no longer periodic (see Fig.~\ref{Fig2}b). The phase $\phi(\bm{r})$ does a full turn as $\bm{r}$ moves around the vortex core, but since $d_0$ remains finite the local gap never closes (see Fig. \ref{fig:sketch} and supplementary video.)  
Such a Kekulé 2D vortex binds a Majorana-like mode at its core at the Dirac point frequency, which in our concrete example of a bolted elastic plate takes the form of a strongly localized vibrational mode of flexural wave motions and out-of-plane bolt vibrations as visualized in Fig.~\ref{fig:sketch}. 

We fabricated such vortex-hosting configuration using the bolted-plate design and performed measurements with a laser Doppler vibrometer (LDV). In Fig.~\ref{Fig2}c, we show the experimental setup, in which a piezoelectric actuator excites the core of the vortex, and the LDV takes point-by-point measurements of the plate to reconstruct the wave field in the entire scanned area (see Methods). 

To numerically characterize the system, we model the bolts as resonators with an effective out-of-plane stiffness and employ the multiple scattering theory  (MST) \cite{torrent2013elastic}. For fixed $n=1$ and $\xi=0.1a$ in Eq. (2), we vary the gap opening perturbation $d_0$, and calculate the local density of states (LDOS) (see Methods for more details). In Fig.~\ref{Fig2}e, we plot the emerging spectral peaks and bandgap region (grey region) as we increase the perturbation $d_0$ in the vortex. For $d_0=0$, we recover the unperturbed honeycomb lattice hosting a double Dirac cone at a Dirac frequency of $8.83$ kHz. A nonzero $d_0$ couples the two valleys, thus gapping the two Dirac cones and lifting their  degeneracy, with a gap that increases with $d_0$. Within this bandgap, several localized states emerge. Figure~\ref{Fig2}d illustrates that particularly one state always remains bound inside the gap, and converges within our precision to the Dirac frequency as the perturbation $d_0$ approaches zero. Since a topological state can be eliminated only by closing a bulk bandgap, we claim that the first peak (blue dots in the figure) corresponds to the topological bound state in our system. The other two states are topologically trivial, since they merge with the continuum below a finite $d_0$, at which point they become delocalized. 

In Fig.~\ref{Fig2}e we plot in blue the $n=1$ normalized LDOS, computed for fixed $d_0=0.15a$, with the topological mode highlighted. To corroborate its topological origin, we also include in red the LDOS results for the zero-winding $n=0$, $\xi \to 0$ configuration of Fig.~\ref{Fig2}a, which should not trap topological states. Its LDOS shows a similar bandgap, but with peaks clustering around the edges of the bandgap.  
We experimentally verify the existence of vortex-localized states by plotting the measured normalized LDOS in Fig.~\ref{Fig2}f. We detect three localized states at frequencies 8.83 kHz (topological), 9.81 kHz, and 10.17 kHz. We attribute the slight frequency mismatch between MST and experiments to fabrication errors and to the fact that the MST model only captures the out-of-plane motion of the bolts \cite{chaunsali2018experimental}. The computed spatial profile of the displacement field at 8.83 kHz, excited by the point-like actuator shown in Fig.~\ref{Fig2}c, indicates that the topological mode is concentrated at the $n=1$ vortex core (Fig.~\ref{Fig2}h), while the displacement field is hardly visible in the $n=0$ non-resonant case (Fig.~\ref{Fig2}g). This prediction is closely matched by the measured mode profile shown in Fig.~\ref{Fig2}i.

\begin{figure*}
   \centering
   \includegraphics[width=0.9\textwidth]{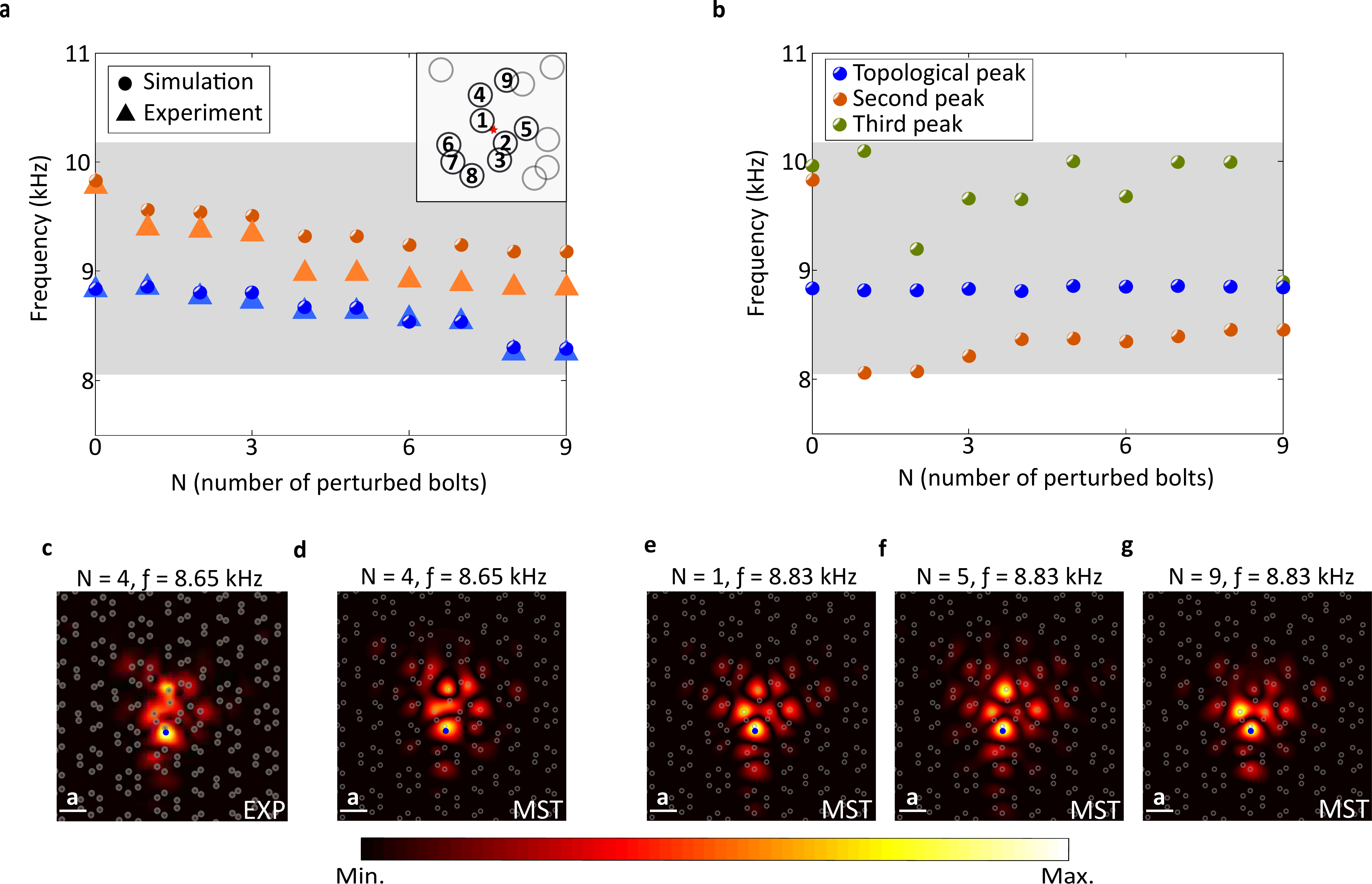}    
   \caption{\textbf{Topological robustness with and without particle-hole symmetry.} \textbf{a}, Dependency of the in-gap bound state frequencies with an increasing number of mass-loaded bolts. \textbf{b}, Bound state frequency evolution under PH-symmetry-preserving perturbations. Note the frequency pinning of the topological mode. Insets show an enlarged view of the vortex core with numbered perturbation sites. \textbf{c,d}, Measured and simulated bound state map when $N=4$ bolts are perturbed through mass-loading. \textbf{e--g}, Simulated, mechanical bound states protected by particle-hole symmetry for $N=1$, $N=5$, and $N=9$, respectively.}
   \label{fig:3}
\end{figure*}

Topological bound states within a given symmetry class are special in that they cannot be removed out of the gap by arbitrary local perturbations that remain in the same symmetry class. To confirm the topological nature of the 8.83kHz mode trapped at the vortex, we now study its behavior as local perturbations are added to the core region. In the first place, we introduce a small local perturbation by adding a mass. This we do by fastening two nuts (with $\sim$41\% mass of each bolt) to $N$ bolts sequentially as indicated in insert of Figs.~\ref{fig:3}a,b. The mass-loading perturbation does not preserve the particle-hole (PH) symmetry of the original symmetry class, and hence the topological bound state is not truly protected against this type of perturbation. As the number $N$ of loaded bolts is increased, all modes shift, including the topological one, and may even move out of the gap.
Before exiting the gap, however, the experimental and numerical displacement maps of the topological mode remain remarkably insensitive to the perturbation, as can be seen by comparing Figs.~\ref{fig:3}c,d to Figs.~\ref{Fig2}h,i. This result highlights the topological mode's exceptional robustness even against symmetry-breaking local mass perturbations. Additional measurements are found in the Extended Data Figs 1 and 2. True topological protection as enjoyed by MBSs is stronger, however, and implies a pinning of the mode to the Dirac frequency as the symmetric perturbation grows. To confirm such topological pinning, we designed a special perturbation that preserves PH symmetry by simultaneously changing the bolt stiffness and mass, see Methods. We show numerically that this perturbation, sequentially added to the same set of bolts, shifts the energy of all modes in the gap except for the topological mode, which indeed remains pinned at 8.83 kHz, see Figure~\ref{fig:3}b. In addition, like in the mass loading experiment, the spatial profile of the non-trivial mode remains essentially insensitive to the local perturbations, see Figs.~\ref{fig:3}e-g. These observations confirm the mode to be a mechanical analogue of a Majorana bound state.

We numerically and experimentally demonstrated that the mechanical analogue of a MBS can exist in artificial structures hosting a non-trivial Kekulé vortex.  We specifically showed the effect on the topologically bound mode of local-mass perturbations and PH symmetry-preserving perturbations. In contrast to other trivial bound states in the system, we showed that the topological mode is completely insensitive to PH-symmetric perturbations, which confirms its unique topological character. We foresee that our findings will widen the research of exotic topological phases in bosonic settings and could stimulate robust control and guiding of mechanical energy for signalling and filtering applications.

\section*{Acknowledgements}
J. C. acknowledges the support from the European Research Council (ERC) through the Starting Grant No. 714577 PHONOMETA and from the MINECO through a Ram\'on y Cajal grant (Grant No. RYC-2015-17156). J. Y. gratefully acknowledges the support from the NSF (CAREER1553202 and EFRI-1741685). P. S-J. acknowledges support from MINECO/FEDER through Grant No. FIS2015-65706-P. N.L and J.V.A.  acknowledge financial support from MINECO grant FIS2015-64886-C5-5-P. D.T. acknowledges financial support through the ``Ram\'on y Cajal'' fellowship under grant number RYC-2016-21188.

\section*{Author Contributions}
J. C. and P.S.-J. conceived and directed this project. J. K. and R. C. guided the experimental work that was conducted by C.-W. C. The numerical computations were conducted by N. L. under the supervision of D. T. and J. V. A. C.-W. C. wrote the article and P.S.-J., J.V.A., N.L. and J.C. undertook revisions. All authors contributed to the discussion and manuscript preparation.

\section*{Competing interests}
The authors declare no competing interests.

\section*{METHODS}
\subsection*{Sample fabrication}
The sample is made of a thin aluminum 6061-T6 plate (762 $\times$ 762 $\times$ 2 mm ) and M4 black-oxide alloy steel bolts. The lattice size $a$ is 26 mm. First, we machined and threaded 1069 holes on the plate using a CNC milling machine. We ensured that the bolts were firmly and equally fastened by taking advantage of the partially threaded bolts, in which the thread-less part can be used as limiter. We then used an electric screwdriver to tighten the bolts with the same torque as much as possible. At the same time, we added an instant-bond adhesive (Loctite® 431) on the threads to secure the contact between the bolts and the plate. For the robustness study, we attached two zinc-plated steel nuts (1.4 g in total) at the bottom of each bolt (3.4 g). 

\subsection*{Experimental measurements and post-processing}
We bonded a piezoelectric ceramic disc actuator (STEMiNC, diameter 10 mm, and thickness 1 mm) with a conductive silver epoxy adhesive just below the center of the plate, at which the topological mode has the maximum out-of-plane displacement. We send a frequency-chirped signal (2--20 kHz in 100 ms) from a function generator to the actuator via a voltage amplifier. We used a laser Doppler vibrometer (Polytec OFV 5000) to detect these vibrations. The device was mounted on a 2-axis linear stage, automated to scan a 2D area on the plate. We conducted point-by-point measurements in a square grid (7.5 $\times$ 7.5 mm) and collected the velocity-time history of 1600 points in total inside the yellow region shown in Fig.~\ref{Fig2}c. All the measurements were synchronized with respect to the onset of the input voltage signal of the function generator. We performed a fast Fourier transformation (FFT) on these velocity-time histories and obtained the power spectral of density (PSD) to reconstruct a 2D field map at a given frequency. We further calculated the LDOS by summing the squared PSD for all points in the scanning area. 

\subsection*{Modelling}
\subsubsection{Multiple scattering theory}
The multiple scattering method is employed to solve the plate biharmonic equation coupled to a cluster of harmonic oscillators. We compute the plate displacement field as a superposition of a known harmonic incident wave $\psi_0(\vec{r},t)=\psi_0(\vec{r})e^{i\omega t}$ and the iterated scattered wave at each resonator\cite{torrent2013elastic}.
 
Each resonator is modelled as a point scatterer. The incident wave is taken as a point source. The plate stiffness, width and density together with the harmonic oscillator masses and spring constants are encoded in two dimensionless parameters, $\Omega_R$ and $\gamma_R$. These two quantities define uniquely the Dirac frequency $\Omega_D$ for the undistorted lattice, see Fig.~\ref{fig:ext1}. In this article we use $\Omega_R=2.15$ and $\gamma=10$. The spatial integration of absolute value of the displacement field is proportional to the local density of states (LDOS).

\subsubsection{PH symmetric bolt-perturbation}
The Dirac frequency in dimensionless units is $\Omega_D=1.362$ which transformed to kHz is $f_D=8.83$kHz. This corresponds to $t=1$ in Extended Data Fig.~\ref{fig:ext1}. The Dirac frequency changes  as a function of either $\gamma$ or $\Omega_R$ . However, an appropriate change of the two keeps the Dirac frequency constant. More precisely,

\begin{equation}
\gamma'=\frac{1}{S_0}\left|\frac{1}{\Omega_D^2}-\frac{1}{\Omega_R'^2}\right|
\label{EqphS}
\end{equation}
where $S_0$ is a constant equivalent to the one deduced in Ref. \cite{torrent2013elastic}.
\newpage

\setcounter{figure}{0}  
\renewcommand{\figurename}{Extended Data Fig.}
\begin{figure*}
\centering
\includegraphics[width=6in]{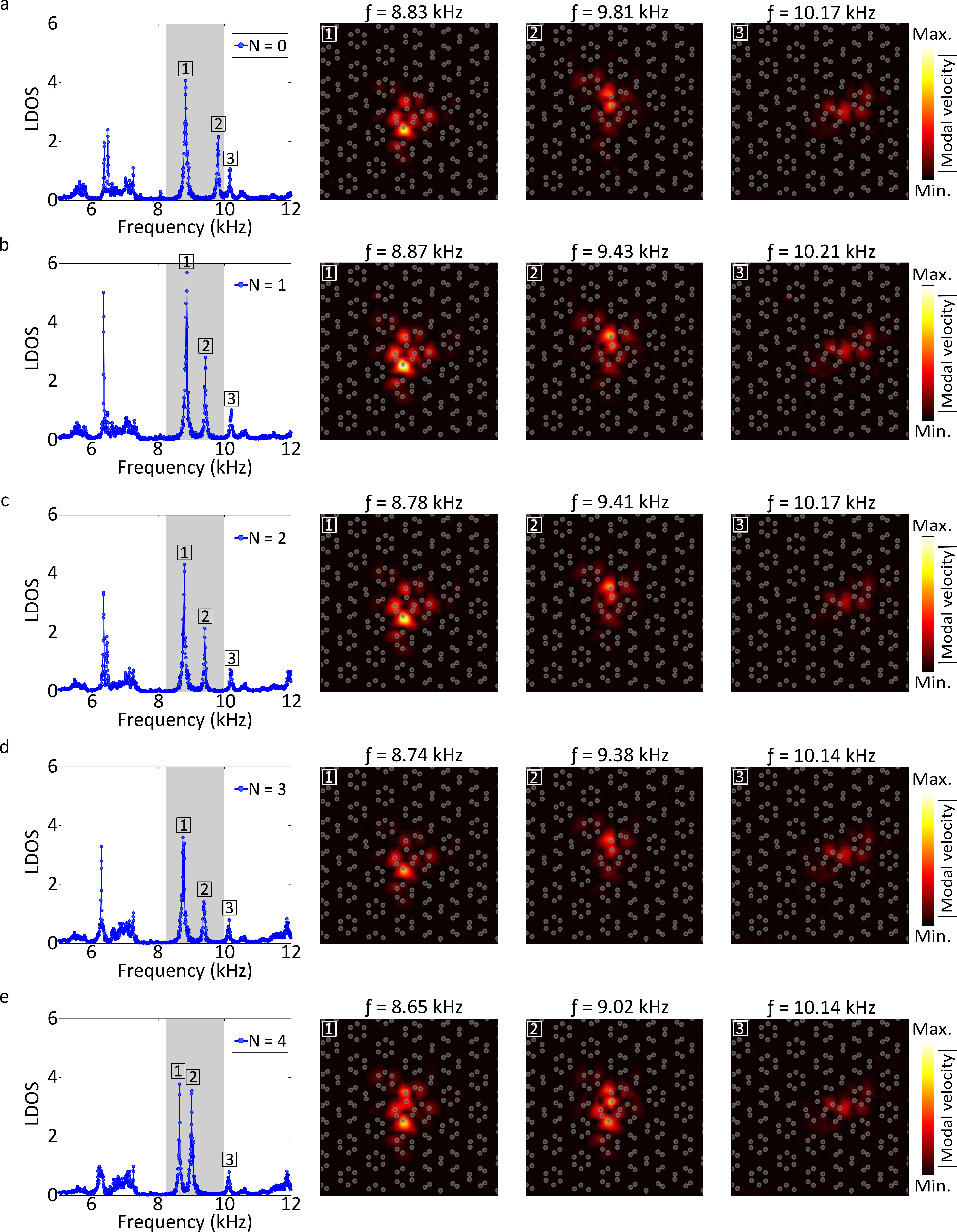}
\caption{\textbf{Experimentally measured LDOS and field maps.} We introduced perturbations in the form of mass-loaded bolts, with \textbf{a}, N=0. \textbf{b}, N=1. \textbf{c}, N=2. \textbf{d}, N=3. \textbf{e}, N=4.}
\label{fig:s2}
\end{figure*} 

\begin{figure*}
\centering
\includegraphics[width=6in]{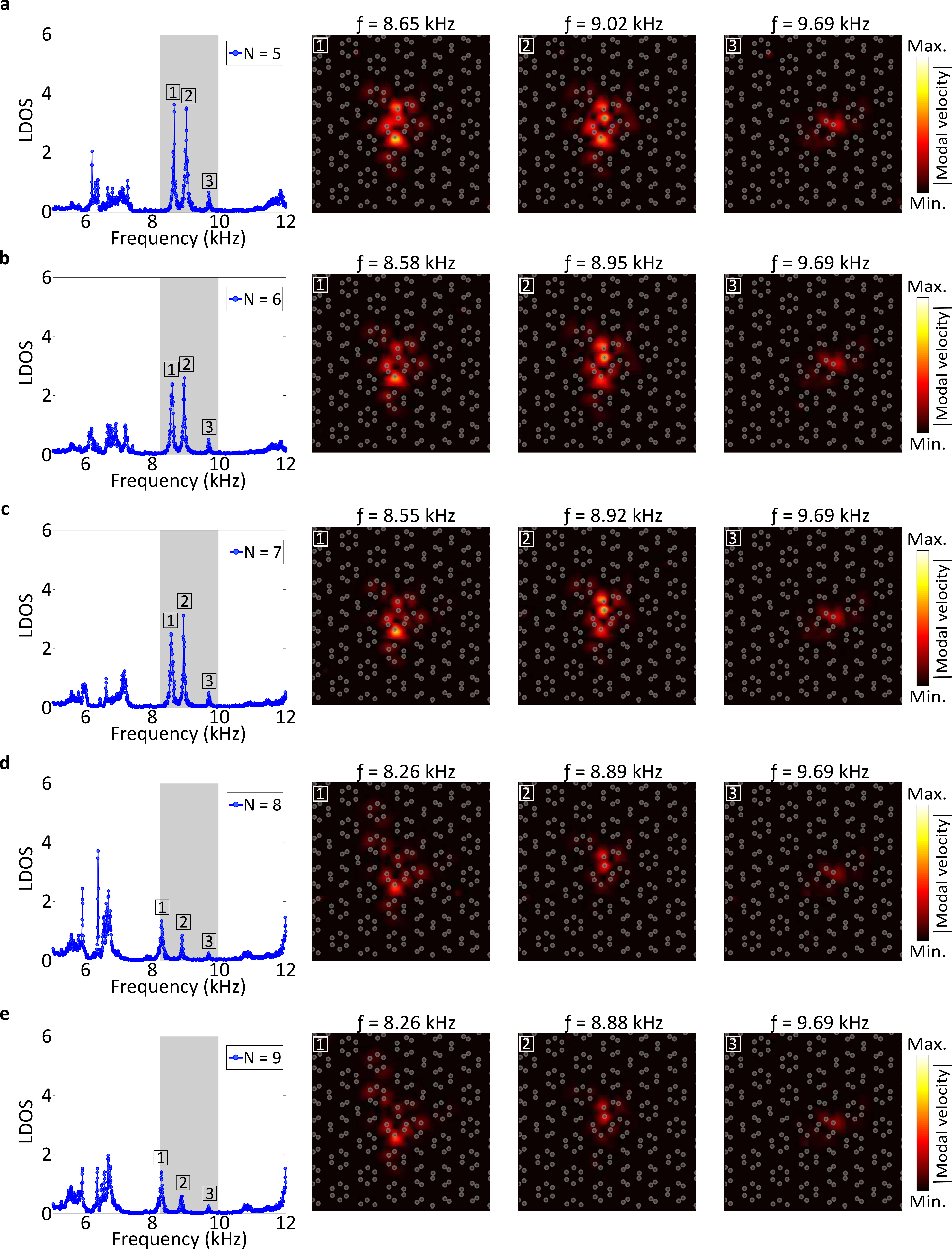}
\caption{\textbf{Experimentally measured LDOS and field maps.} We introduced perturbations in the form of mass-loaded bolts, with \textbf{a}, N=5. \textbf{b}, N=6. \textbf{c}, N=7. \textbf{d}, N=8. \textbf{e}, N=9.}
\label{fig:s3}
\end{figure*} 

\begin{figure*}
\centering
   \includegraphics[width=0.70\textwidth]{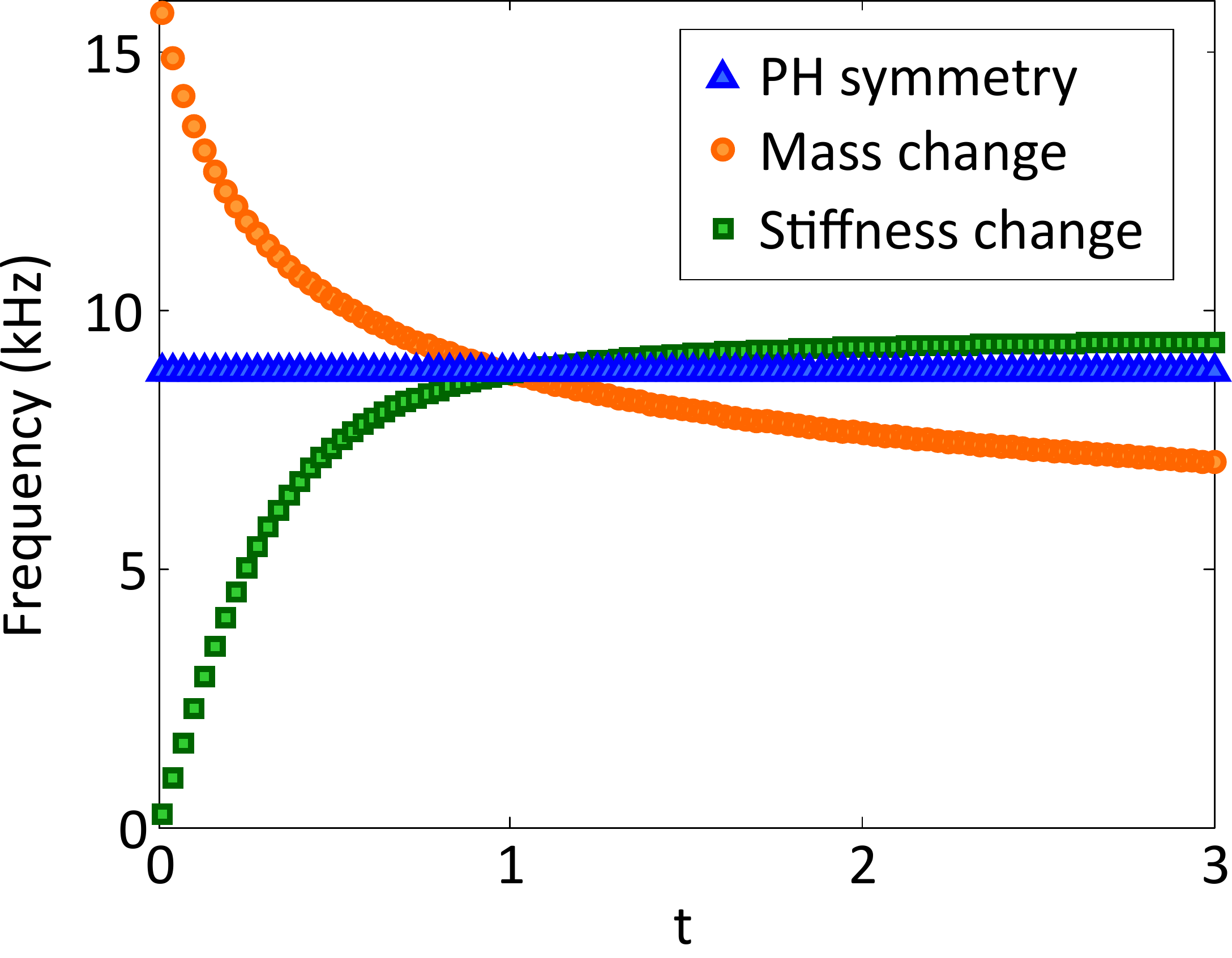}
\caption{\textbf{Spectral location of Dirac frequency $f_D$ as a function of $t$.} The mass change is computed for fixed $\Omega_R$ and varying $\gamma'=\gamma t$. The stiffness change correspond to fixed $\gamma$ and $\Omega_R'=\Omega_R t$. The PH symmetric corresponds to varying both parameters simultaneously according to $\gamma'(\Omega_R')$ in Eq. \eqref{EqphS}. }
\label{fig:ext1}
\end{figure*} 

\end{document}